\documentclass[fleqn,usenatbib]{mnras}

\usepackage{newtxtext,newtxmath}

\usepackage[T1]{fontenc}
\usepackage{ae,aecompl}

\usepackage{graphicx}	
\usepackage{amsmath}	
\usepackage{amssymb}	
\usepackage{url}
\usepackage{enumitem}

\usepackage{boldline}

\title[Dust Self-Scattering in HD 163296]
{Validating Scattering-Induced (Sub)millimeter Disk Polarization through the Spectral Index, Wavelength-Dependent Polarization Pattern, and Polarization Spectrum: The Case of HD 163296}

\author[Z. D. Lin et al.]{
Zhe-Yu Daniel Lin$^{1}$\thanks{E-mail: zdl3gk@virginia.edu}, 
Zhi-Yun Li$^{1}$, 
Haifeng Yang$^{2}$, 
Leslie Looney$^{3}$,
\newauthor
Ian Stephens$^{4}$, 
and Charles L. H. Hull$^{5,6,7}$
\\
$^{1}$Department of Astronomy, University of Virginia, 530 McCormick Rd., Charlottesville 22904, Virginia, USA\\
$^{2}$Institute for Advanced Study, Tsinghua University, Beijing, 100084, People's Republic of China \\
$^{3}$Department of Astronomy, University of Illinois at Urbana-Champaign, Urbana IL 61801, USA \\
$^{4}$Harvard-Smithsonian Center for Astrophysics, 60 Garden Street, Cambridge, MA 02138, USA \\
$^{5}$National Astronomical Observatory of Japan, NAOJ Chile Observatory, Alonso de C\'{o}rdova 3788, Vitacura, Santiago, Chile \\
$^{6}$Joint ALMA Observatory, Alonso de C\'ordova 3107, Vitacura, Santiago, Chile \\
$^{7}$NAOJ Fellow
}

\date{Accepted XXX. Received YYY; in original form ZZZ}

\pubyear{2019}

\begin{document}
\label{firstpage}
\pagerange{\pageref{firstpage}--\pageref{lastpage}}
\maketitle

\begin{abstract}

A number of young circumstellar disks show strikingly ordered (sub)millimeter polarization orientations along the minor axis, which is strong evidence for polarization due to scattering by $\sim0.1$ mm-sized grains. To test this mechanism further, we model the ALMA dust continuum and polarization data of HD 163296 using RADMC-3D. We find that scattering by grains with a maximum size of 90 $\mu$m simultaneously reproduces the polarization observed at Band 7 and the unusually low spectral index ($\alpha \sim 1.5$) between Band 7 and Band 6 in the optically thick inner disk as a result of more efficient scattering at the shorter wavelength. The low spectral index of $\sim 2.5$ inferred for the optically thin gaps is reproduced by the same grains, as a result of telescope beam averaging of the gaps (with an intrinsic $\alpha \sim 4$) and their adjacent optically thick rings (where $\alpha \lesssim 2$). The tension between the grain sizes inferred from polarization and spectral index disappears because the low $\alpha$ values do not require large mm-sized grains. In addition, the polarization fraction has a unique azimuthal variation: higher along the major axis than the minor axis in the gaps, but vice versa in the rings. We find a rapidly declining polarization spectrum (with $p \propto \lambda^{-3}$ approximately) in the gaps, which becomes flattened or even inverted towards short wavelengths in the optically thick rings. These contrasting behaviors in the rings and gaps provide further tests for scattering-induced polarization via resolved multi-wavelength observations. 

\end{abstract}




\section{Introduction}

(Sub)millimeter polarization observations of disks around young stellar objects have been increasing in number, spatial resolution, and wavelength coverage \citep[e.g.][]{Rao2014, Stephens2014,Cox2015,Kataoka2016, Fernandez2016, Stephens2017, Alves2018, Bacciotti2018, Harris2018, Hull2018,  LeeCF2018, Girart2018, Dent2019, Harrison2019, Mori2019, Sadavoy2018_VLA1632, Sadavoy2018, Sadavoy2019, Vlemmings2019}. The original scientific goal was to determine the magnetic field in disks through magnetically aligned grains, analogous to the interpretations of large scale star-forming regions \cite[e.g.][]{Girart2006, Stephens2013, Hull2014}. The magnetic fields are widely believed to play a key role in disk structure and dynamics \citep[e.g.][]{Blandford1982, Balbus1991, Turner2014} which determine the conditions of planet formation \citep[e.g.][]{Morbidelli2016}.

However, the theoretical interpretations of polarization observations of disks remain uncertain. In addition to alignment with the magnetic field via radiative torques \citep[e.g.][]{Andersson2015}, there are several other mechanisms that can align non-spherical grains, such as through radiative anisotropy \citep{Lazarian2007, Tazaki2017}, mechanical alignment torques \citep{Lazarian2007_MAT, Kataoka2019}, or aerodynamic alignment \citep{Gold1952, Lazarian1995}. Furthermore, the temperature structure may influence the interpretation of the alignment direction when disks are optically thick \citep{Yang2017_nearfar, Ko2019, Lin2020} and so does the porosity of non-spherical grains as long as the grain size is larger or comparable to wavelength \citep{Kirchschlager2019}. To further complicate the situation, spherical grains can also produce polarization by dust self-scattering \citep{Kataoka2015, Yang2016_HLTau, Yang2017_nearfar}. Polarization by self-scattering depends on the radiation field of the observing wavelength. For inclined axisymmetric disks, the observational features can be simple: the polarization orientation tends to be along the disk minor axis; the level of polarization is roughly $\sim 1 \%$; and the level of polarization is generally expected to decrease with increasing wavelength in the limit of scattering by grains smaller than the observing wavelength in an optically thin medium. Indeed the scattering interpretation of disk polarization is favored in several targets \citep[e.g][]{Stephens2014, Cox2015, Kataoka2016,Stephens2017,Bacciotti2018,Hull2018,LeeCF2018,Girart2018,Harris2018,Dent2019}. However, given the plethora of polarization mechanisms, the scattering interpretation may not be unique. Independent evidence for dust scattering is highly desirable to validate the interpretation.

One way to validate scattering at (sub)millimeter wavelengths is through the dust continuum spectral index, $\alpha$, defined by $\alpha \equiv d \log I_{\nu} / d\log \nu$ where $I_{\nu}$ is the intensity. Consider an isothermal slab of purely emitting dust (without scattering) with an optical depth of $\tau$. In the Rayleigh-Jeans and optically thin ($\tau \rightarrow 0$) limit, $\alpha = \beta_{\text{dust}} + 2$ where $\beta_{\text{dust}}$ is the dust opacity index, defined by $\kappa_{\text{abs}} \propto \nu^{\beta_{\text{dust}}}$. $\kappa_{\text{abs}}$ is the dust absorption opacity. It is expected that $0 < \beta_{\text{dust}} \lesssim 2$ depending on the composition or size of the grain \citep[e.g.][]{Draine2006}. In the optically thick limit ($\tau \rightarrow \infty$), $\alpha = 2$, which is the frequency dependence of the Rayleigh-Jeans Law, and the information of $\beta_{\text{dust}}$ is lost. From an observational standpoint, it is convenient to define an inferred opacity index by $\beta \equiv \alpha - 2$. In this case, $\beta = \beta_{\text{dust}}$ only in the optically thin limit ($\tau \rightarrow 0$) and $\beta =0$ in the optically thick limit ($\tau \rightarrow \infty$). When scattering is included, \cite{Birnstiel2018} demonstrated that an increase in the albedo of grains decreases the emerging intensity, i.e., scattering makes objects appear dimmer. \cite{Zhu2019} further investigated the wavelength dependence of the one-dimensional scattering slab and deduced that, in the optically thin limit, $\alpha = \beta_{\text{dust}} +2$, which is the same relation as that without scattering. The difference is in the optically thick limit, where the albedo determines the spectral index: spectral index can be lower than 2 for an albedo that decreases with increasing wavelength. \cite{Liu2019} demonstrated that the low spectral index of TW Hya can be explained by grains with maximum sizes on the order of 10-100 $\mu$m when scattering is included. This maximum grain size is on the scale of the grains inferred from polarization of other targets such as HD 142527 \citep{Kataoka2016}, HL Tau \citep{Kataoka2016_HLTau}, and IM Lup \citep{Hull2018}. If scattering is the favored mechanism for the polarization and the spectral index can in principle be explained by scattering, it is natural to ask: can scattering simultaneously reproduce both the polarization fraction and the spectral index in a particular object quantitatively? If so, the case for scattering would be greatly strengthened.

We seek to address the above question for the well-studied disk of the Herbig Ae star HD 163296. It is a Class II source at a distance of 101 pc \citep{Gaia2018}. Rings and gaps in the disk were first resolved in \cite{Isella2016} using the Atacama Large Millimeter/Submillimeter Array (ALMA) and further substructures were resolved in The Disk Substructures at High Angular Resolution Project \citep[DSHARP;][]{Andrews2018, Isella2018}. The disk is a particularly suitable source to study polarization and the spectral index since both measurements are available \citep{Dent2019}. It was shown in \cite{Dent2019} that the spatially resolved polarization image can be explained by dust self-scattering \citep[see also][]{Ohashi2019}. In particular, the polarization orientation is mainly along the disk minor axis and the anticorrelation of polarization fraction to the total intensity (Stokes $I$) of the dust emission along the major axis is broadly consistent with the expected anisotropy of the radiation field that plays an important role in the scattering-induced polarization \citep{Kataoka2015, Yang2016_HLTau}. However, the model first explored in \cite{Dent2019} did not reproduce the low spectral index even though scattering was included. In particular, the central region of the disk, within a radius of $\sim 40$AU, has $\alpha < 2$ (i.e., $\beta < 0$). The unique combination of polarization pattern and low spectral index of HD 163296 provides a rare opportunity to quantitatively test whether dust scattering can reproduce both effects.

The structure of the paper is as follows: we prescribe a disk model in Section \ref{sec:modelsetup} and calculate the expected images based on a Monte Carlo radiative transfer code. The results are presented in Section \ref{sec:Results}. We discuss the results in Section \ref{sec:discussion} and conclude in Section \ref{sec:conclusion}. 

\section{Model Setup} \label{sec:modelsetup}
In this section, we describe the dust opacity model and disk model considered to study HD 163926. To compare with observations, we will use the Monte Carlo radiative transfer code RADMC-3D \footnote{\url{http://www.ita.uni-heidelberg.de/~dullemond/software/radmc-3d/}} to produce observable quantities including polarization. The independent observational constraints considered are the high-angular resolution observations at Band 6 ($\lambda$ = 1.25mm) from DSHARP\footnote{\url{https://bulk.cv.nrao.edu/almadata/lp/DSHARP/}}\citep{Andrews2018, Huang2018, Isella2018} with a synthesized beam FWHM of 0\farcs038 $\times$ 0\farcs048, the polarization fraction at Band 7 ($\lambda$ = 0.87mm) with a FWHM of $0\farcs21 \times 0\farcs19$, and the spectral index with a FWHM of $0\farcs18 \times 0\farcs17$ presented in \cite{Dent2019}.

\subsection{Dust Opacity} \label{ssec:opacity}
We follow the DSHARP dust model prescribed in \cite{Birnstiel2018}. Opacity is calculated using Mie Theory which assumes compact (non-porous) spherical grains and requires as input parameters the grain size and refractive indices. The dust model is a mixture of water ice, astronomical silicates, troilite, and refractory organic material with mass fractions of $\sim 0.2$, $0.33$, $0.07$, and $0.4$, respectively. The respective refractive indices of the different components are taken from \cite{Warren2008}, \cite{Draine2003}, and, for the latter two, \cite{Henning1996}. Furthermore, we average over a grain size distribution that follows a simple power-law, $a^{-3.5}$, with minimum and maximum grain size cut-offs \citep[the standard MRN grain size distribution; ][]{Mathis1977}, where $a$ is the grain size. The minimum grain size is fixed at $0.01\mu$m, while the maximum grain size $a_{\text{max}}$ is left as a free parameter. The precise value for the minimum grain size makes little difference at the considered wavelengths \citep{Kataoka2015, Birnstiel2018}. Mie opacity calculation is implemented using the python version of the Mie code available in RADMC-3D. The Mie code was ported from the Fortran 77 version by Bruce Draine \footnote{\url{https://www.astro.princeton.edu/~draine/scattering.html}} that is based on the original code published by \cite{Bohren1983}. 

\begin{figure}
    \centering
    \includegraphics[width=\columnwidth]{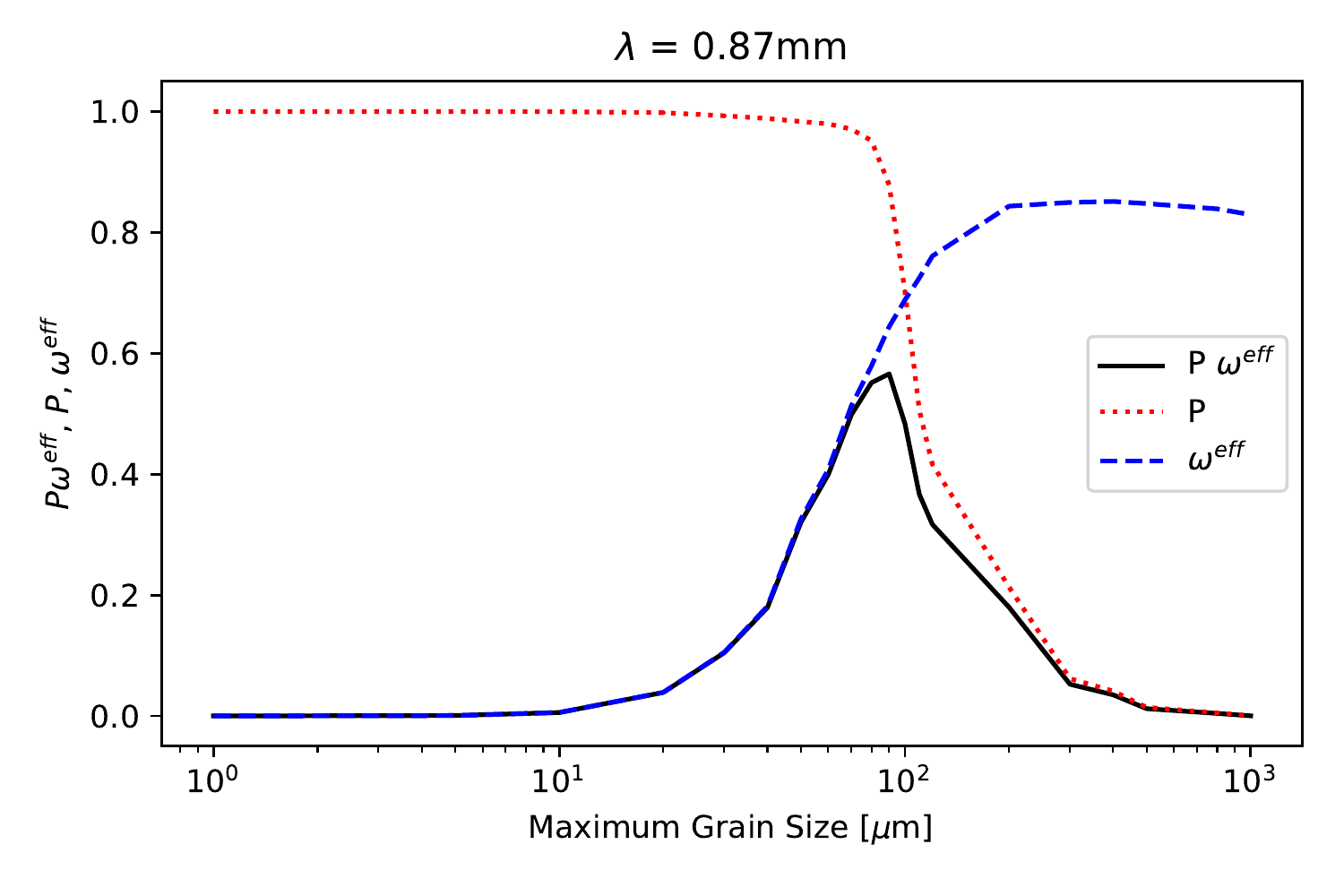}
    \caption{
        The calculated degree of linear polarization for single scattering ($P$; red dotted line), effective albedo discussed in Section \ref{ssec:opacity} ($\omega^{\text{eff}}$; blue dashed line), and the product $P \omega^{\text{eff}}$ (black solid line) at $\lambda=0.87$mm for various grain sizes. 
    }
    \label{fig:grainsize4pol}
\end{figure}

A useful quantity to assess the dust scattering-produced polarization at a given wavelength, $\lambda$, is the product of the single scattering albedo, $\omega_{\nu}$, and single scattering polarization fraction taken at $90^{\circ}$, $P=-z_{12}/z_{11}$, where $z_{12}$ and $z_{11}$ are elements of the scattering matrix \citep{Kataoka2015}. $\omega_{\nu}$ increases rapidly as the grain size parameter $x \equiv 2 \pi a/\lambda$ approaches unity, but $P$ drops quickly for large ($x > 1$) grains. These two competing effects mean that the polarization at a given wavelength is determined by grains of a certain range of sizes \citep{Kataoka2015}. For a particle with a size larger than the wavelength, forward scattering dominates $\omega_{\nu}$ even though forward scattering is effectively not scattering. A simple replacement for the apparent scattering efficiency is the effective albedo, $\omega^{\text{eff}}_{\nu}=(1-g_{\nu})\omega_{\nu}$ where $g_{\nu}$ is the forward-scattering parameter \citep{Ishimaru1978, Birnstiel2018, Tazaki2019, Zhu2019}. Fig. \ref{fig:grainsize4pol} shows the calculated $P $ and $\omega_{\text{eff}}$ for ALMA Band 7 ($\lambda$ = 0.87mm) as a function of the maximum grain size. The product, $P \omega^{\text{eff}}_{\nu}$, gives the grain size that contributes most to the polarization at the observing wavelength. Fig. \ref{fig:grainsize4pol} shows that the optimal maximum grain size for producing polarization at $\lambda$ = 0.87mm is $90 \mu$m. We will therefore hold $a_{\text{max}}$ fixed at this value for our models and assume the same dust population throughout the whole disk; note that this value is somewhat smaller than the fiducial grain size of $140 \mu$m adopted by \cite{Ohashi2019} for their independent modeling of the HD 163296 disk polarization. The corresponding absorption opacity, effective albedo, and the forward-scattering parameter are $\kappa_{\text{abs}}=0.97$ cm$^{2}$ g$^{-1}$ per gram of dust, $\omega_{\nu}^{\text{eff}} = 0.64$, and $g_{\nu} = 0.15$ at $\lambda$ = 0.87mm, whereas at $\lambda=$1.25mm, $\kappa_{\text{abs}} = 0.46$ cm$^{2}$ g$^{-1}$, $\omega_{\nu}^{\text{eff}} = 0.47$, and $g_{\nu} = 0.044$. The opacity index $\beta_{\text{dust}}$ index between the two wavelengths is 2.06. 

\subsection{Disk Model} \label{ssec:diskmodel}

We consider a dust density distribution that is determined by distributions of surface density and scale height. We follow the DSHARP notation in \cite{Huang2018} to describe the substructures of HD 163296. For example, ``B100" represents the bright annular feature with a radius of 100 au. To model HD 163296-like disks with prominent substructures (rings and gaps), we adopt a composite surface density distribution $\Sigma_{d}$ that is composed of two different types of profiles: power-law-like profiles and Gaussian-like profiles. The former consists of a power-law distribution within a region between radii of $R_a$ and $R_b$ and an exponential decrease interior and exterior to the region:

\begin{equation}
    \Sigma_{p}(R) = \Sigma_{p,0} \times \begin{cases}
                    \exp{ 
                        \bigg[ -\dfrac{1}{2} 
                                \bigg( \dfrac{R - R_{a}}{\sigma_{a}} \bigg)^{2} 
                        \bigg]
                        }            \text{ , } R < R_{a} \\
                    (
                        \dfrac{R}{R_{a}}
                        )^{-p}   \text{ , } R_{a} \leq R < R_{b}\\
                    (
                        \dfrac{R_{b}}{R_{a}}
                        )^{-p} 
                        \exp{ 
                            \bigg[ -\dfrac{1}{2}\bigg( \dfrac{R-R_{b}}{\sigma_{b}}\bigg)^{2} 
                            \bigg]
                            }    \text{ , } R \geq R_{b}
            \end{cases}
    \text{ ,}
\end{equation}
where $R$ is the cylindrical radius,  $\Sigma_{p,0}$ is the surface density at the inner edge $R_{a}$ of the range and $p$ is the power-law index. The quantities $\sigma_{a}$ and $\sigma_{b}$ correspond to the characteristic distance over which the surface density decreases interior to $R_a$ and exterior to $R_b$, respectively. This type of profile is particularly suitable for modeling the inner disk ($R<50$AU) and the outer B100 region of HD 163296. Since gaps may not be entirely free of material, we adopt a surface density floor using the same power-law-like profile but with $p=0$. We do not consider the innermost ring (B14) and gap (D10), because those features are not resolved in the spectral index map or in the polarization image that we seek to model. We also ignore the small-scale non-axisymmetric (crescent-like) feature interior to B67 for simplicity. 

The Gaussian-like profile is prescribed as 
\begin{equation}
    \Sigma_{G}(R) = \Sigma_{G,0} \times \begin{cases}
                    \exp{ \bigg[-\dfrac{1}{2}\bigg( \dfrac{R-R_{0}}{\sigma_{c}}\bigg)^{2} \bigg]} & \text{ , } R \le R_{0} \\
                    \exp{ \bigg[-\dfrac{1}{2}\bigg( \dfrac{R-R_{0}}{\sigma_{d}}\bigg)^{2} \bigg]} & \text{ , } R > R_{0}
                \end{cases}
    \text{ ,}
\end{equation}
where $R_{0}$ is the radius of the center of the profile, $\Sigma_{G,0}$ is the surface density at $R_{0}$, and $\sigma_{c}$ and $\sigma_{d}$ are the widths of the Gaussian distributions interior and exterior to $R_{0}$ respectively. This type of profile is particularly suited for modeling rings. The full surface density profile is obtained by summing all the various power-law-like and Gaussian-like profiles:

\begin{equation}
    \Sigma_{d} = \sum_{i} \big( \Sigma_{p} \big)_{i} + \sum_{i} \big( \Sigma_{G} \big)_{i} \text{ ,}
\end{equation}
where $i$ is the dummy variable that iterates through the number of assigned components. 

For the temperature distribution of the large (90 $\mu$m) grains that are responsible for the scattering-induced polarization in our model, we adopt the following simple power-law: 
\begin{equation}
    T = T_{t} \bigg( \dfrac{R}{R_{t}}\bigg)^{-q} \text{ ,}
\end{equation}
where $T_{t}$ is the temperature at a characteristic radius $R_{t}$ and $q$ is the power-law index\footnote{We refrain from introducing additional parameters to characterize the vertical variation of the temperature of the dust, because, as we show later, the scale height of the approximately 0.1~mm-sized grains in our model is significantly smaller than that of the gas, which should make the temperature in the relatively thin dust layer similar to that of the midplane for a passive disk.}. For our Monte Carlo radiative transfer calculations of the disk polarization and spectral index, the mass density of the dust must be specified. It is parameterized by 
\begin{equation}
    \rho(R,z) = \dfrac{\Sigma_d (R)}{\sqrt{2\pi} H} \exp{ \bigg[ -\dfrac{1}{2}\bigg( \dfrac{z}{H}\bigg)^{2} \bigg]} \text{ ,}
\end{equation}
where $z$ is the cylindrical height and $H$ is the dust scale height, parameterized as $H=H_{t} (R/R_{t})^{q/2 + 3/2}$. Ideally, both the dust and temperature distributions should be computed self-consistently, but doing so is numerically costly and it would require additional assumptions about the level of turbulence in the disk and the abundance of small grains which absorb the stellar radiation near the disk surface. These are beyond the scope of this paper.

\subsection{Using the 1D Slab Model to Estimate Parameters} \label{ssec:slabmodel}
To constrain model parameters, one can implement a grid-based search or other parameter fitting methods, e.g., Markov-Chain Monte Carlo. However, polarization radiative transfer calculations are computationally expensive; thus we choose to first estimate the parameters needed to fit the Bands 6 and 7 data observed in HD 163296 at several representative locations with the help of a one-dimensional (1D, plane-parallel) analytic model.

Following \cite{Birnstiel2018} and \cite{Zhu2019}, we consider a 1D isothermal slab with scattering where the observed intensity at a given wavelength can be computed analytically for different optical depths and temperatures. We fix the inclination angle to  $46.7^{\circ}$, which was measured by fitting ellipses to the continuum annular substructures \citep{Isella2018}. Using the opacity determined in Section \ref{ssec:opacity}, we search through a parameter space of temperature and surface density of the slab and calculate the expected intensities at ALMA Bands 6 and 7 and the spectral index between the two. 

\begin{figure*}
    \centering
    \includegraphics[width=\textwidth]{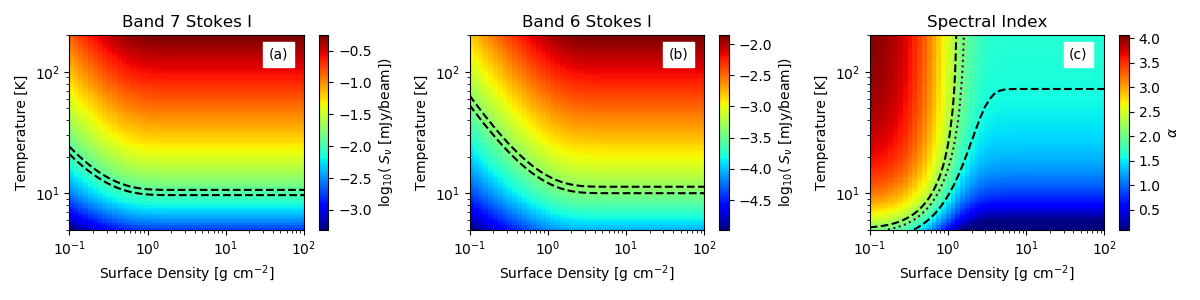}
    \caption{
            The plane-parallel calculation of an isothermal scattering slab in the two-dimensional parameter space of surface density and temperature. The panels from the left to right are Band 7 Stokes $I$, Band 6 Stokes $I$, and spectral index between the two ALMA bands. The dashed lines bracket the observed range for ring B100 of HD 163296. The dotted line in panel c marks where $\alpha = 2$. 
        }
    \label{fig:exmp_param}
\end{figure*}

The color maps of Fig. \ref{fig:exmp_param} show the calculated intensity at Bands 7 and 6 and the spectral index. The intensities are scaled by the respective finite beam sizes of the observations for direct comparison in the remainder of this section. For a given temperature, the intensity increases with increasing surface density and reaches an asymptotic value as the slab becomes optically thick (see Equation 12 of \citealt{Zhu2019}) as seen in Fig. \ref{fig:exmp_param}a and \ref{fig:exmp_param}b. From Fig. \ref{fig:exmp_param}c, we see that the spectral index in the optically thin limit measures the opacity index determined by the absorption opacity and decreases to a constant value determined by the albedo and viewing inclination as the optical depth increases (see Equations 21 and 22 of \citealt{Zhu2019}). The constant value in the optically thick regions can have $\alpha < 2$, i.e., $\beta < 0$, (the right side of the dotted line in Fig. 2c). We can also see that at low temperatures where the Rayleigh-Jeans limit is inapplicable, the spectral index decreases faster and also reaches a lower value. This is because the Planck function produces lower intensity at higher frequency than that in the Rayleigh-Jeans limit and thus creates a lower spectral index. 

We directly compare with the observations using the values at B100 as a demonstration in Fig. \ref{fig:exmp_param}. We overlay the range that encloses $10\%$ of the measured Stokes $I$ at Bands 7 and 6 to account for the absolute flux calibration uncertainty of ALMA observations and an uncertainty of 0.3 for the measured spectral index (\citealt{Dent2019}; which used a $7\%$ calibration uncertainty). There is a limited range of surface density and temperature that can simultaneously reproduce the data. These constrained ranges serve as a consistency check and provide a starting point for exploring the parameter space using Monte Carlo radiation transfer calculations in Section \ref{sec:Results}. 

\begin{figure*}
    \centering
    \includegraphics[width=\textwidth]{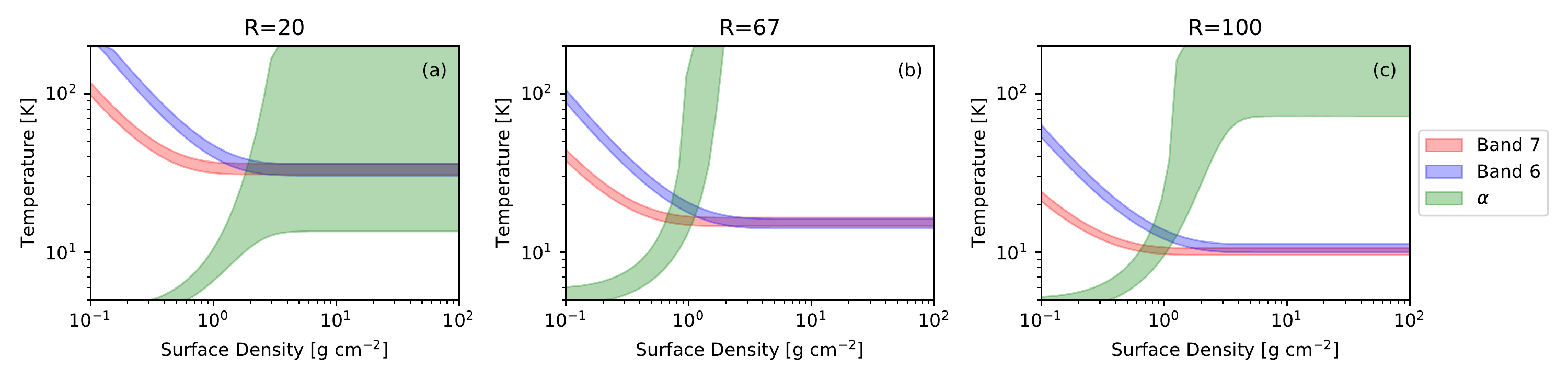}
    \caption{
            The range of surface density and temperature constrained by observations at 20, 67 and 100 au. The red, blue, and green regions correspond to ranges constrained by the Band 7 Stokes $I$, the Band 6 Stokes $I$, and the spectral index between Band 7 and Band 6. 
        }
    \label{fig:param_range}
\end{figure*}

In Fig. \ref{fig:param_range}, we overplot the regions that fulfill the observed constraints within its uncertainty for the three representative radii of the bright bands: 20, 67 and 100 au. The constraints are listed in Table \ref{tab:constraints}. In panel \ref{fig:param_range}a, there is a threshold surface density above which increased surface density does not alter the intensity and spectral index. This threshold is achieved simply because both wavelengths become optically thick, thus the intensity only depends on wavelength or temperature. At B67, or panel \ref{fig:param_range}b, there is only a limited range of surface densities suggesting that the ring cannot be optically thin or very optically thick (unlike the inner disk). For both panels \ref{fig:param_range}b and c, the intersecting range is in the region where spectral index falls quickly with surface density, which hints that these rings are not in the Rayleigh-Jeans limits. 

\begin{table}
    \centering
    \begin{tabular}{c c c c}
        Radius [au] & Band 7 [mJy beam$^{-1}$] & Band 6 [mJy beam$^{-1}$] & $\alpha$ \\
        \hline
        20 & 78.1 & 2.08 & 1.4\\
        67 & 26.2 & 0.76 & 2.2 \\
        100 & 12.0 & 0.44 & 1.9 
    \end{tabular}
    \caption{
            Values from observations used for 1D slab estimation in Section \ref{ssec:slabmodel}. The first column is the radius in au. The second to fourth columns are the observed values for the fluxes in Bands 7 and 6 and the spectral index $\alpha$ between the two bands. 
        }
    \label{tab:constraints}
\end{table}

\section{Numerical Results} \label{sec:Results}

\begin{table}
    \centering
    \begin{tabular}{l c c c c}
        \multicolumn{5}{c}{Parameters} \\
        \hline
        $R_{t}$ [au] & \multicolumn{4}{c}{100} \\
        $T_{t}$ [K] & \multicolumn{4}{c}{12.5} \\
        $q$ & \multicolumn{4}{c}{0.65} \\
        $H_{t}$ [au] & \multicolumn{4}{c}{3} \\
        
        \hlineB{2}\\
    
        \multicolumn{5}{c}{Power-law-like Surface Density} \\
        \hline \\
        Feature             & Inner Disk    & D48   & D86   & B100 \\
        \hline \\
        $\Sigma_{p,0}$ [g cm$^{-2}$]  & 8             & 0.02     & 0.1     & 0.5 \\
        $p$                 & 0             & 0     & 0     & 7    \\
        $R_{a}$ [au]        & 0.5             & 38     & 76     & 105 \\
        $R_{b}$ [au]        & 26             & 58     & 96     & 135\\
        $\sigma_{a}$ [au]   & 0             & 5     & 5     & 5 \\
        $\sigma_{b}$ [au]   & 6             & 4.5     & 5     & 10 \\
        
        \hlineB{2} \\
        \multicolumn{5}{c}{Gaussian-like Surface Density} \\
        \hline \\
        Feature             & B67           & B100      & B155\\
        \hline \\
        $\Sigma_{G,0}$ [g cm$^{-2}$]     & 2.5      & 2.1   & 0.15\\
        $R_{0}$ [au]        & 67            & 100   & 155\\
        $\sigma_{c}$ [au]   & 5             & 3     & 5 \\
        $\sigma_{d}$ [au]   & 5             & 4     & 25 \\
        \hline
        
    \end{tabular}
    \caption{
            A chosen set of parameters for the model. The first column displays the name of the parameter from Section \ref{sec:modelsetup} and its corresponding units in squared brackets. The parameters are separated into three main groups: the general parameters, parameters used for power-law-like surface density components, and parameters used for Gaussian-like surface density components. For the inner disk, the inner exponential decrease is ignored since it is not resolved and thus $\sigma_{a}=0$. 
        }
    \label{tab:fittedparameters}
\end{table}

\subsection{Polarization Image} 

\begin{figure*}
    \centering
    \includegraphics[width=\textwidth]{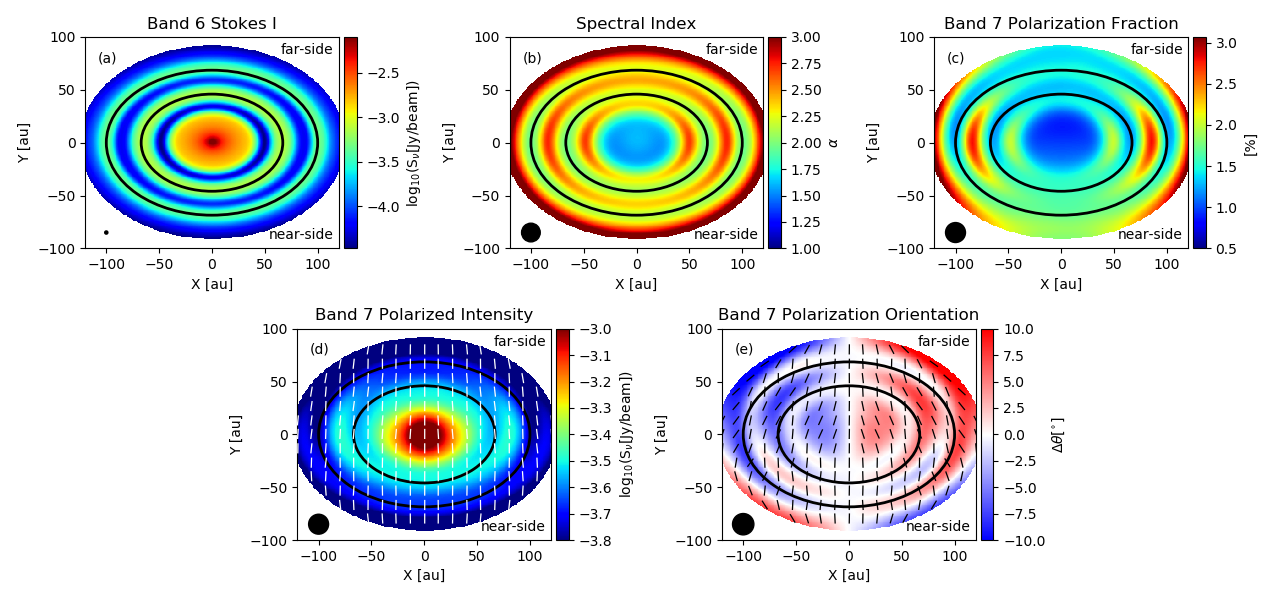}
    \caption{
        The two dimensional images of the disk model. The near-side of the disk is the bottom half of the image. Panel a: the Band 6 Stokes $I$ image in Jy beam$^{-1}$. Panel b: the spectral index between Band 7 and Band 6. Panel c: the Band 7 polarization fraction in percent. Panel d: the Band 7 polarized intensity in Jy beam$^{-1}$, with the polarization vectors overplotted. Panel e: the color scale depicts the offset of the polarization orientation, $\Delta \theta$, from the vertical direction (parallel to the disk minor axis). $\Delta \theta$ is plotted east from north (counter-clockwise) and $\Delta \theta=0$ means the polarization orientation is parallel to the disk minor axis. The vectors in Panel e are oriented from the vertical direction by $3 \times \Delta \theta$ to make the offsets obvious. The black ellipses for all six panels denote the rings. The Gaussian beams used for convolution are plotted as black filled circles at the lower left.
            }
    \label{fig:polimage}
\end{figure*}

Guided by the analytical estimates in Section \ref{ssec:slabmodel}, we performed a set of RADMC-3D-based radiative transfer calculations including polarization from scattering. A chosen set of parameters for the model that fits the observational data for HD 163296 is presented in Table \ref{tab:fittedparameters}. We plot the image results in Fig. \ref{fig:polimage}. To consider finite resolution, the Stokes $I$ images are convolved by a Gaussian beam with a circular FWHM of 0\farcs043 for Band 6. The polarized intensity, polarization fraction, and polarization orientation at Band 7 are produced after the Stokes $Q$ and $U$ images are first convolved individually with a circular FWHM of 0\farcs2. The spectral index is calculated after Band 7 and Band 6 Stokes $I$ are each convolved by a Gaussian beam with a circular FWHM of 0\farcs175 first to match that from \cite{Dent2019}. The values for the FWHM are the geometric mean of the major and minor axes of the respective synthesized beams \footnote{We use only circular FWHM beams for simplicity; nevertheless, the major and minor axes of the spectral index map are very similar. The Band 6 DSHARP beam is more elliptical, although the precise beam shape should be important only for the small structures which were not resolved in Band 7. We have checked that using the elliptical beam size for the Band 6 DSHARP data does not significantly modify the results.}. The images are only plotted up to 125au in radius because we are mainly interested in the region with detected polarization \citep{Dent2019}.

One can quickly verify that the Stokes $I$ of Band 6 is qualitatively similar to the observations. The inner disk and the rings produce higher intensity relative to the gaps. Furthermore, the general polarization features at Band 7 match those of the observations. The polarized intensity is larger in the inner disk and rings. The polarization orientations are mainly in the disk minor axis direction, within $\sim\pm 10^{\circ}$ across the image. We plot the deviation, $\Delta\theta$, from the disk minor axis ($x=0$) in panel (e) of Fig. \ref{fig:polimage} in order to show the variation of polarization orientation. 

The distribution of polarization orientation is similar to that observed and modeled by \cite{Dent2019} and \cite{Ohashi2019}. The polarization orientation is largely along the disk minor axis, which is expected for a disk \citep{Yang2016_HLTau, Yang2017_nearfar}. The distribution of the deviation reflects the radiation anisotropy previously discussed in \cite{Kataoka2015, Yang2016_HLTau, Yang2017_nearfar}. For example, in an optically thin gap, most of the radiation comes from the adjacent rings or the inner disk, which is in the radial direction. The radial radiation anisotropy leads to a slightly more azimuthal polarization orientation in the gaps, as observed (see Fig.~1d of \citealt{Dent2019}). 

\subsection{Profiles along the Major Axis}

\begin{figure*}
    \centering
    \includegraphics[width=\textwidth]{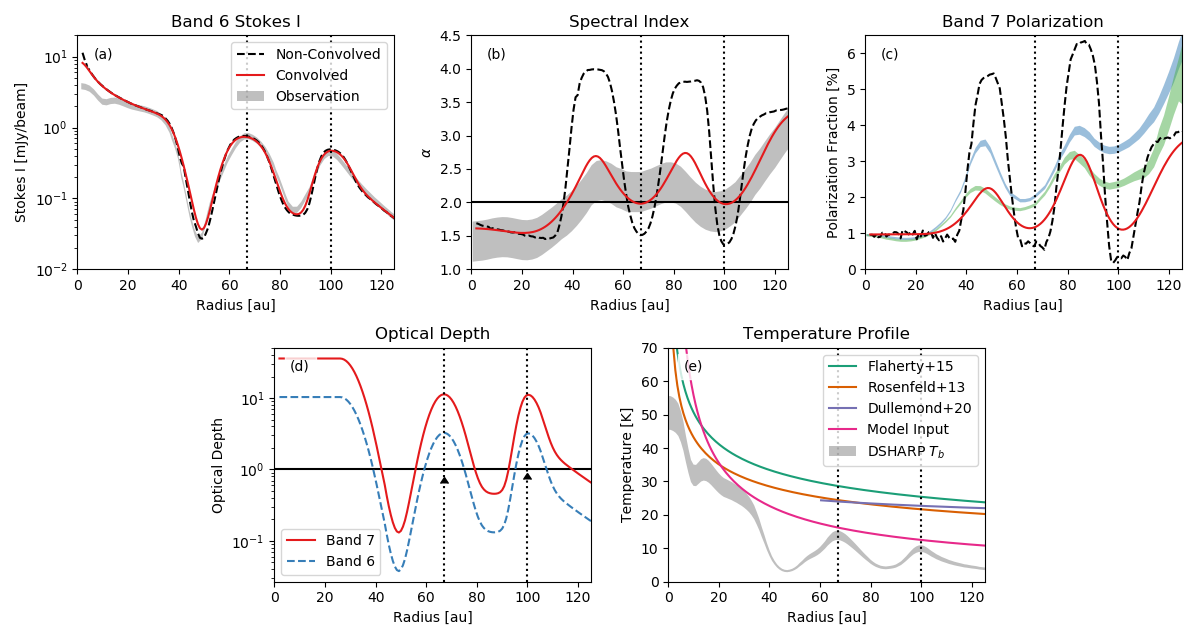}
    \caption{
        Panel a: the 0\farcs043 resolution Band 6 Stokes $I$ image in mJy per beam. The units of the non-convolved image is scaled to the corresponding mJy per beam. The convolved profile is plotted with a red solid line, the non-convolved profile with a black dashed line, and the grey region represents the $\pm 10\%$ calibration uncertainty of the data. Panel b: the spectral index between Band 7 and Band 6. The grey region is $\alpha \pm 0.3$, where $\alpha$ is the spectral index averaged between the Northwest and Southeast major axes. The $\alpha=2$ is plotted as a horizontal solid line. Panel c: we plot the convolved Band 7 polarization with a red solid line and the non-convolved polarization with a black dashed line. The green and blue colored regions are the observed Band 7 polarization with uncertainty from random noise and image fidelity along the Northwest and Southeast major axes of the disk \citep{Dent2019}. Panel d: the model optical depth along the major axis for Bands 7 and 6. The lower limits for Band 6 optical depths at 67 and 100 au (denoted by triangles) are derived from the level of $^{12}$CO extinction through the dust rings \citep{Isella2018}. The optical depth of 1 is plotted as a horizontal solid line. Panel e: a comparison of various dust temperature profiles as a function of radius from previous literature. The vertical dotted lines through all the panels are the locations of the rings. 
            }
    \label{fig:obsmaj}
\end{figure*}

For more quantitative comparisons, we take a cut along the major axis and compare the beam-convolved model images to observations in panels a to c of Fig. \ref{fig:obsmaj} which show, respectively, the Stokes $I$ of Band 6, the polarization fraction in Band 7, and the spectral index between Bands 7 and 6. The polarization data is taken from Fig. 2a of \cite{Dent2019} where the polarization along the major axis is separated into the Northwest and Southeast directions. Note that the low spectral index values of $\sim 1.5$ and the polarization fraction of $\sim 1 \%$ in the inner disk are well reproduced. The model optical depths along the major axis at Band 7 and 6 are plotted in panel d.

To identify the effects of convolution, we also plot the intrinsic model images without convolution. In the gap regions, the intrinsic spectral index approaches $\alpha \sim 4$, which is expected since the grains we are using have $\beta_\text{dust} = 2.06$ (recall in the Rayleigh Jeans limit, $\alpha = \beta_\text{dust} + 2$). In the optically thick rings, the spectral indices are intrinsically low ($\sim 1.5$) because of preferential scattering at shorter wavelengths, although deviation of the black body radiation from the Rayleigh-Jeans limit also plays a role, especially for the outer, colder ring. The observed rings have $\alpha \sim 2$, which is not as low as that in the inner disk since the emission from the rings is blended with emission from the gaps that have much higher spectral indices.

The polarization fraction after convolution coincides with the observed polarization within the first gap, but is somewhat lower than the observed values at outer radii given by \cite{Dent2019}. However, our model matches better the polarization fraction presented in \cite{Ohashi2019} at outer radii (see, e.g., the top panel of their Fig. 18). We should caution the reader that the polarization fraction may be significantly affected by the differential filtering of Stokes $I$ vs. $Q$ and $U$ by interferometers such as ALMA. This effect has yet to be thoroughly quantified, and will be explored elsewhere.


\section{Discussion} \label{sec:discussion}

\subsection{Consistency Check and Discrepancies}

\subsubsection{Optical Depth}
We find that a grain size of $90 \mu$m, similar to the $100 \mu$m found in \cite{Dent2019}, reproduces their observed level of polarization in the inner disk, although it is lower beyond the first gap. However, the spectral index they obtained was $\sim3.5$ even in the inner disk and rings, which is most likely due to a lower optical depth adopted in their model compared with ours. For the inner disk at radius $R\lesssim 30$ au, we find that optical depths of $\sim10$ at Band 6 and higher for Band 7 are needed to produce the low spectral index observed there. The high optical depth of the inner disk implies that one would not be able to see the line emission from the back side of the disk \citep{Zhu2019}. This seems to be the case judging from the DSHARP $^{12}$CO channel maps \citep{Isella2018}; however, confirmation of this claim is beyond the scope of this paper. For the dust rings at 67 and 100 au, we find optical depths of $\sim 3$ at their centers in Band 6. These values are consistent with the constraints provided by \cite{Isella2018}, who found lower limits of $\sim 0.6$ and $0.7$, respectively, based the level of extinction of the $^{12}$CO emission from the back side of the disk. 

\subsubsection{Temperature}

In Fig. \ref{fig:obsmaj}e, we compare the brightness temperature at Band 6 and the input temperature profile as a function of radius. Our temperature determination is based on the spectral index between Band 6 and 7 and on the Band 7 polarization data. It is particularly driven by the temperature at the outer, 100 au ring. The fact that this ring has a spectral index $\alpha$ close to 2 (see the last column of Table \ref{tab:fittedparameters} and Fig.~\ref{fig:obsmaj}b) indicates that it is most likely optically thick (this is discussed further in Appendix \ref{app:Tmid}), unless the grains responsible for most of its emission are much larger than the observing wavelengths. However, such large grains would not scatter the photons at the observing wavelengths efficiently and would have difficulty reproducing the observed polarization \citep{Kataoka2015, Yang2016_HLTau}. If the outer ring is indeed optically thick, its temperature would naturally be close to the relatively low brightness temperature that we observe (shown in Fig.~\ref{fig:obsmaj}b as shaded line), which is indeed what we find from our modeling; the small difference between the two is attributed to scattering \citep{Birnstiel2018}. We therefore believe that, broadly speaking, our midplane dust temperature makes physical sense.

In Fig. \ref{fig:obsmaj}e, we further plot the midplane temperature profiles found from existing literature based on CO lines, but scaled to our adopted distance of 101pc: $19 (R/128 \text{au})^{-0.3}$K \citep{Rosenfeld2013}, $22.5 (R/124 \text{au})^{-0.3}$K \citep{Flaherty2015}, and $18.7 (R / 400 \text{au})^{-0.14}$K for $R>60$au \citep{Dullemond2020}. We find that our midplane dust temperature is lower than these profiles. In particular, \cite{Dullemond2020} presented a clever way of inferring the gas temperature at the disk midplane based on the brightness temperature of optically thick CO lines at the line of sight velocity expected of the midplane. As stressed by \cite{Dullemond2020}, this method relies on the assumption that CO molecules are present at the midplane and are not frozen out onto dust grains. Their inferred gas temperature drops from $\sim$25K at $\sim$100 au to $\sim$18K at $\sim$400 au, although there are significant variations, especially near the 100 au ring, where T$\sim$20K. In the inner, $\sim$100 au region that is of direct interest to us, their inferred temperature is close to but above the CO freeze-out temperature. If optically thick CO is present at the midplane, the gas (and dust) temperature at the midplane would indeed be of order 20 K or more, which would be significantly higher than the dust temperature that we infer at the same radius. However, if the CO is frozen-out at the midplane, it would not probe the midplane temperature. It would probe, instead, the temperature of the so-called ``CO snow surface" above (or below) the midplane where the warmer CO-rich gas transitions to the colder CO-poor gas (and dust). If the snow surface is spatially close enough to the midplane, it may not be easy to kinematically distinguish it from the midplane, especially in view of the fact that the CO lines are broadened by thermal motions, optical depth effects, and potentially some level of turbulence \citep{Flaherty2015, Flaherty2017}. While the results of \cite{Dullemond2020} by themselves are consistent with a relatively warm midplane temperature of 20 K or more on the 100 au scale, they do not exclude the possibility of a colder midplane that is out of the reach of their CO-based method because of CO freeze-out. Therefore, we conclude that our dust-based temperature does not necessarily contradict those CO-based temperature estimates from the literature, particularly that of \cite{Dullemond2020}.

In fact, if the dust temperature at the midplane were as high as those inferred from optically thick CO lines, we would have difficulty reproducing the continuum data. This is illustrated in Appendix \ref{app:Tmid}, where we redid the model shown in Fig.~\ref{fig:obsmaj} using a higher temperature profile from \cite{Flaherty2015}. The higher temperature model does not fit the spectral index, especially at the outer, 100 au ring. As stated earlier, this is simply because the higher temperature model requires less dust (and thus a lower dust optical depth) but the $\sim 100\ \mu$m-sized grains needed for scattering-induced polarization cannot produce the observed flat spectral index of $\alpha\sim 2$ unless they are optically thick. This is strong evidence that the disk midplane is colder than the temperatures inferred from optically thick CO lines, which most likely probe the warmer gas from the disk surface down to the CO snow surface rather than the colder midplane region, as discussed above.

The Band 6 Stokes $I$ for the innermost part of the disk ($\lesssim 10$~au) appears to be higher than the observed value, indicating that the simple power-law temperature profile breaks down at small radii. The prescribed temperature profile also does not take into account the rings and gaps, which may be required for more precise fitting.

\subsubsection{Dust Scale Height}

In our model, the dust scale height is found to be 3 au at a radius of 100 au. For a disk with dust settling, the dust vertical distribution is balanced by the downward gravity and the upward turbulent mixing. This results in a height that is a fraction of the gas pressure scale height \citep{Dubrulle1995, Armitage2015}. For an estimation, we approximate the fraction by:
\begin{align}
    \dfrac{H_{d}}{H_{g}} \sim \sqrt{\dfrac{\alpha_{v}}{\Omega \tau_{f}}} \text{ ,}
\end{align}
where $H_{d}$ is the dust scale height and $H_{g}$ is the gas pressure scale height. The quantity $\alpha_{v}$ here is the viscosity parameter (not to be confused with the spectral index); $\Omega$ is the Keplerian orbital frequency; and $\tau_{f}$ is the frictional timescale defined by $\tau_{f} \equiv \rho_{s} a / \rho_{g} c_{s}$ where $\rho_{s}$ is the mass density of the solid, $\rho_{g}$ is the gas density, and $c_{s}$ is the thermal sound speed. We estimate the gas density structure from the multi-line fitting results in \cite{Flaherty2015} scaled to our adopted distance of 101pc. At a radius of 100 au, $H_{g} = 8.7$au, gas surface density is 3.9 g cm$^{-2}$ and $\rho_{g} = 1.2 \times 10^{-14}$ g cm$^{-3}$. From Section \ref{ssec:opacity}, the grain size is 90 $\mu$ m and $\rho_{s} \sim 1.7 $ g cm$^{-3}$. We use the temperature found in Section \ref{sec:Results} for $c_{s}$. Using our prescribed dust scale height, the $\alpha_{v} \sim 1.7 \times 10^{-3}$. The upper limit on the viscosity parameter was found to be $\alpha _{v} \lesssim 1 \times 10^{-3}$ \citep{Flaherty2015} which is broadly consistent with our model result. \cite{Ohashi2019} considered two different dust populations with different scale heights, $H_{d}$, at different radii to fit the polarization and found that $H_{d} / H_{g} \lesssim 1/3$ for $R \lesssim 70$au and $H_{d} / H_{g} \sim 2/3$ for $R \gtrsim 70$au. Our results are consistent with the conclusion from \cite{Ohashi2019} and using a larger dust scale height at larger radii may increase the polarization fraction for our model.

\subsection{Apparent Low Opacity Index from Beam Averaging of Rings and Gaps}
\label{subsec:beam}


Measurements of the spectral index at (sub)millimeter wavelengths have long been used to infer grain sizes in protoplanetary disks. In particular, a relatively low inferred opacity index $\beta \equiv \alpha -2 \lesssim 1$ has often been taken as evidence for grains up to millimeter size or larger \citep{Beckwith1991, Testi2014}. Such large (mm) sizes are in tension with the grain sizes inferred from disk polarization if the polarization arises from dust scattering \citep{Kataoka2015}. For example, in the case of HL Tau, polarization suggested $\sim 100 \mu$m grains \citep{Yang2016_HLTau, Kataoka2016_HLTau}, while the apparently low opacity index is taken as evidence for millimeter grains (\citealt{Carrasco2019}, see below).

The same tension appears to exist for the target of the current study, HD 163296: namely, the Band 7 polarization is most naturally produced by $100\mu$m-sized grains through scattering but the apparent opacity index $\beta$ is less than unity at small radii (within about 100~au; see Fig.~\ref{fig:obsmaj}b) which, in the usual interpretation, requires mm-sized grains. However, in this particular case, we have demonstrated that the low values of $\beta$ can be produced by small, $100\mu$m-sized grains because of high-contrast disk substructures (specifically rings and gaps) {\it and} the averaging of the emission from these substructures. In particular, in our best-fit model shown in Fig.~\ref{fig:obsmaj}b, we have $\beta\sim 2$ in the optically thin gaps and $\beta \sim -0.5$ in the optically thick rings. The negative $\beta$ comes from more efficient scattering at the shorter wavelength. After convolution with the telescope beam, the former is brought down to $\beta\sim 0.5$, and the latter up to $\beta \sim 0$. Therefore, in our interpretation, the low apparent opacity index $\beta$ in the inner gaps of the HD 163296 disk does not correspond to $\beta_{\text{dust}}$. As such, it cannot be used to infer the presence of large (mm-sized) grains. In this particular case, the tension in the estimation of grain size from polarization and opacity index $\beta_{\text{dust}}$ may be illusory. An implication is that the grain growth may be stifled around a relatively small size of order $100\mu$m for some reason (e.g., coating of grains by relatively non-sticky CO$_2$ ice that makes it harder for the grains to grow through collisions, \citealt{Okuzumi2019}). Another possibility is that there are indeed larger grains, although it may be hidden within the optically thick $100\mu$m grains.

Our modeling of the specific case of HD 163296 highlights the general need for caution in evaluating the spectral index in disks that have intrinsically high contrast substructures right next to one another, such as rings and gaps. This point was made by \cite{almapartnership2015}, among others, but it is worth stressing, particularly in the context of the aforementioned tension in two different methods of grain size estimation. It is especially true when optically thin and optically thick substructures are located close together and the grains responsible for the emission are much smaller than the observing wavelengths. In such cases, there is rapid spatial variation of the spectral index from $\alpha\sim 4$ in the optically thin substructures to $\alpha \sim 2$ (or less if scattering is important and more efficient at a shorter wavelength) in the optically thick substructures. If the substructures are not (or are only barely) resolved observationally, their fluxes will be blended together, which will artificially increase the spectral index in the optically thick substructures and decrease that in the optically thin substructures. Because of their intrinsically low values, the fluxes in the optically thin substructures are especially susceptible to telescope beam averaging; the substructures must be well resolved in order to prevent the (high) fluxes from the optically thick substructures from leaking into the optically thin regions and overwhelming the (weak) intrinsic emission there. If there is significant contamination from the optically thick substructures, the optically thin regions will have a spectral index substantially lower than $\sim 4$, which may be misinterpreted as evidence for large grains. Conversely, contamination from the optically thin substructures may change the spectral index in an optically thick region from $ \alpha < 2$ to $\gtrsim 2$, thus potentially hiding evidence for strong scattering. 

Given the wide-spread substructures detected in protoplanetary disks through DSHARP \citep{Andrews2018} and other programs, there is a need to re-evaluate whether the relatively low opacity index commonly inferred for disks in the literature \citep{Testi2014}, especially from early low-resolution observations \citep[e.g.][]{Beckwith1991}, is significantly affected by beam averaging or not. If so, it would go a long way towards alleviating the tension between the grain sizes inferred from spectral index studies versus that from scattering-induced polarization studies. High resolution observations that {\it well} resolve the disk substructures at multiple wavelengths are needed to determine whether the tension is real or merely a reflection of scattering and beam averaging.





\subsection{Polarization Spectrum as a Test of Scattering-Induced Polarization}

Besides low ($\alpha \lesssim 2$) spectral index, an additional test of the scattering interpretation of the polarization detected in HD 163296 can come from multi-wavelength data. In the optically thin gaps, the relatively small dust grains that we used in our models to maximize the polarization at ALMA Band 7 ($\lambda$=0.87 mm) are expected to produce lower polarization fractions at longer wavelength bands, since the scattering should be in the Rayleigh regime where the scattering cross-section drops rapidly with increasing wavelength. If the inner disk and the first two rings are really optically thick, as indicated by our modeling, then the polarization spectrum (the distribution of the polarization fraction as a function of wavelength) would be different. In particular, \cite{Yang2017_nearfar} showed that scattering-induced polarization fraction peaks around an optical depth of order unity for an inclined slab, dropping quickly to zero as the optical depth goes to zero and asymptoting to a finite value as the optical depth goes to infinity (see their Fig.~3). It is unclear at the present time whether these expectations are met or not, since there is only polarization data at one Band (Band 7). Here we will make predictions for all other bands where polarization observations are currently possible with ALMA, including Bands 6, 5, 4 and 3 (corresponding to $\lambda=$1.25, 1.5, 2.1, and 3.1 mm, respectively). These can be checked against future observations.

\begin{figure*}
    \centering
    \includegraphics[width=\textwidth]{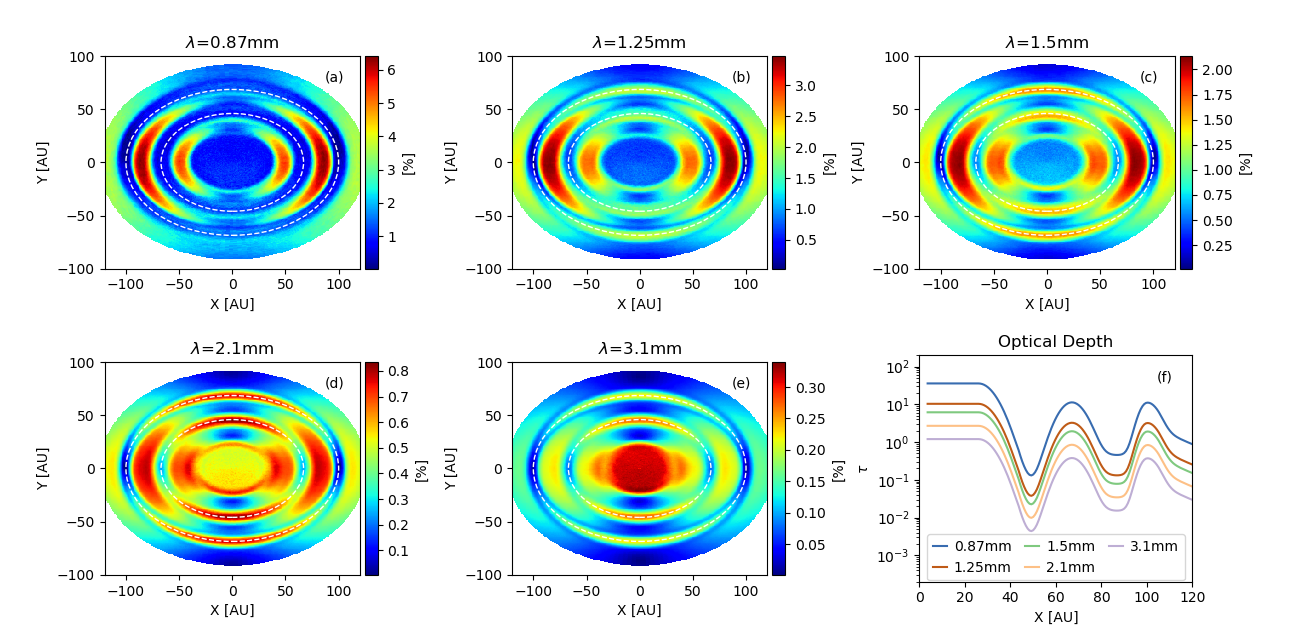}
    \caption{
        Images of the polarization fraction without finite telescope resolution for 5 ALMA bands at $\lambda= 0.87, 1.25, 1.5, 2.1,$ and $3.1$ mm (panels a-e) where the polarization capability is available. The dashed white ellipses mark the location of the rings. Panel f plots the distribution of the optical depth as a function of radius along the major axis for each band. 
            }
    \label{fig:longwavpol}
\end{figure*}

In Fig.~\ref{fig:longwavpol}, we plot the spatial distribution of the polarization fraction in each of the ALMA bands in color scale (panels a-e) without effects of finite telescope spatial resolution. We also plot the radial distribution of the optical depth along the major axis for each band (panel f). Note that the maximum value for the color scale is different for each band, starting at $\sim 6\%$ for the shortest-wavelength Band 7 and decreasing to $\sim 0.35\%$ for the longest-wavelength Band 3. The change of polarization fraction with wavelength is discussed in detail below. Here, we focus on the spatial distribution of the polarization fraction, which has an interesting transition from Band 7 to Band 3. For all images, the azimuthal variation along the rings versus that along the gaps are opposite. Along the rings, the polarization fraction is higher along the minor axis ($x=0$). Along the gaps, the polarization fraction is higher along the major axis ($y=0$). The former trend is most clearly seen in the Band 3 image, while the latter trend is most obvious at Band 7.

The azimuthal variation can be understood in terms of two competing effects: the inclination-induced polarization and the radiation-anisotropy-induced polarization. For both the rings and gaps, the inclination-induced polarization produces polarization parallel to the disk minor axis, but the radiation anisotropy is different between the rings and gaps. The radiation in the gaps comes mostly from the adjacent rings (or the inner disk), with an anisotropy in the radial direction leading to an azimuthally oriented polarization that adds to the inclination-induced polarization at locations along the disk major axis, but offsets the polarization at locations along the minor axis. On the other hand, the radiation in the rings comes mostly from the azimuthal direction, leading to a radially oriented polarization that offsets the inclination-induced polarization at locations along the major axis, but adds to the polarization at locations along the minor axis \citep[see also][their Fig. 7]{Pohl2016}. Such azimuthal variations for rings and gaps are present at the same time for Band 4 (panel c) and Band 5 (panel d).

Another interesting trend is that, as the wavelength increases, the inner disk becomes more prominent in the polarization fraction map: compared with the rings and gaps at large radii, its polarization fraction is lower at Band 7, but becomes higher at Band 3. The former is due to high optical depth (see panel f), while the latter comes about because the inner disk has an optical depth of order unity at Band 3, which is optimal for producing polarization through scattering \citep{Yang2017_nearfar}. In all cases, the polarization orientations are broadly similar to that displayed in Fig.~\ref{fig:polimage}e for Band 7, with the deviation angles from the disk minor axis increasing somewhat with wavelength. 

\begin{figure}
    \centering
    \includegraphics[width=\columnwidth]{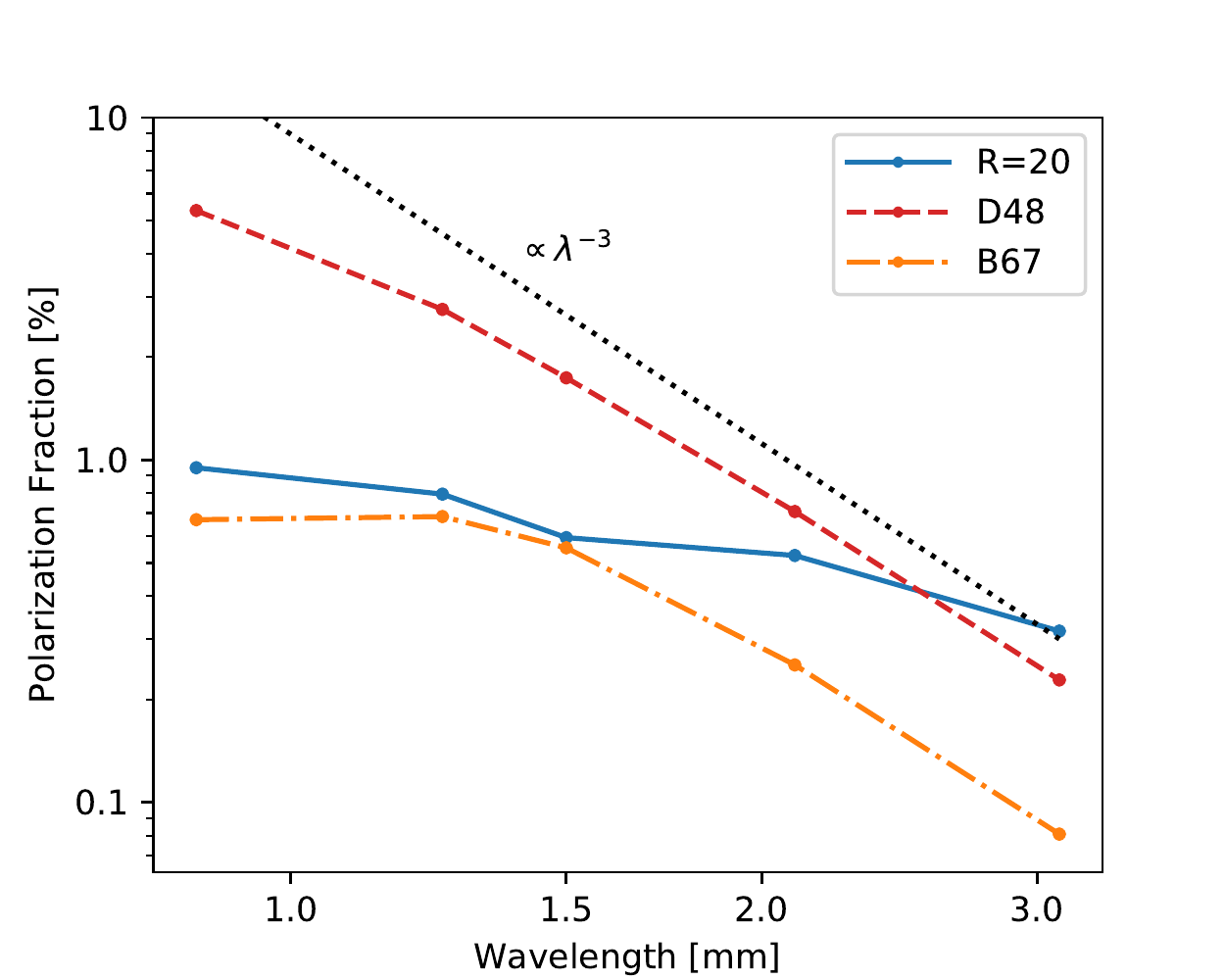}
    \caption{
        Predicted polarization spectra for the inner disk at 20 au (R=20; blue solid line), the D48 gap (red dashed line), and the B67 ring (orange dash-dotted line) along the major axis. The black dotted line shows a power-law that is proportional to $\lambda^{-3}$. 
        }
    \label{fig:pol_spec}
\end{figure}

In Fig.~\ref{fig:pol_spec}, the predicted polarization spectra (polarization fraction as a function of wavelength) are plotted for three representative locations on the major axis: the inner disk at $R=20$au, the D48 gap, and the B67 ring. In the gap, the polarization fraction decreases monotonically with increasing wavelength and approaches the steep power-law $\lambda^{-3}$ for Rayleigh scattering toward the longest wavelength bands with the lowest optical depths. In the ring, the spectrum is drastically different. The polarization fraction is higher for the longer wavelength Band 6 than the shorter-wavelength Band 7, which is the opposite of the common expectation based on the simplest Rayleigh scattering. The reason for this spectral inversion is that the ring is optically thick at both wavebands, but more so at the shorter wavelength band (see Fig.~\ref{fig:longwavpol}f), which reduces the polarization fraction at Band 7 relative to that at Band 6 \citep{Yang2017_nearfar}. This difference in the shape of the polarization spectrum is characteristic of the scattering-induced polarization in highly structured disks such as HD 163296, and should be searched for through multi-wavelength observations. As the wavelength increases further, the ring starts to become optically thin, leading to a drop of the  polarization fraction with wavelength, as expected. The inner disk differs from the ring and gap in that it remains optically thick even at the longest wavelength band. Its polarization fraction decreases with wavelength, but rather slowly, approximately as $\lambda^{-0.85}$, which is much shallower than the naive expectation based on Rayleigh scattering. The shallow spectrum is a result of decreasing optical depth with wavelength, which makes the inner disk detectable in polarization at all 5 ALMA bands. The predicted spectrum for the inner disk, if confirmed by future observations, would provide strong evidence supporting the notion that dust scattering is responsible for both polarization and the low spectral index in the region.

\subsection{Implications for Other Disks}

As mentioned in the introduction, there is evidence that dust scattering is responsible, at least in part, for the (sub)millimeter polarization detected in an increasing number of disks, with HD 163296, IM Lup, and HL Tau as arguably the best examples. Like HD 163296, IM Lup has polarization data published only in one ALMA band \citep[Band 7,][]{Hull2018}, although data have been taken in Band 6 (ALMA project 2019.1.00035.S, PI: C. Hull). Preliminary results indicate that the Band 6 polarization is also oriented along the disk minor axis (as in Band 7) and that its polarization fraction is higher than that in Band 7 (Hull et al., in prep.). This ``inverted" polarization spectrum (compared with the simple expectation based on Rayleigh scattering) could be indicative of the IM Lup disk being optically thick at both Bands 6 and 7, but more so at Band 7 than Band 6, as illustrated by the B67 curve in Fig.~\ref{fig:pol_spec}. There is an indication that the spectral index $\alpha$ between Band 6 \citep{Cleeves2016, Andrews2018} and Band 7 \citep{Cleeves2016,Hull2018} is less than 2 in the central region of the disk (within about 50~au), which is independent evidence for strong dust scattering, as is the case for HD 163296. 

The HL Tau disk has the most complete published multi-wavelength polarization data and spectral index information to date \citep{Kataoka2017,Stephens2017,Carrasco2019}. 
In particular, the disk also has rings and gaps and there exists a similar anti-correlation between the Stokes $I$ and spectral index, which suggests varying optical depth as a function of radius, similar to the case of HD 163296. The polarization at Band 7 is mainly along the disk minor axis and thus favors the dust self-scattering explanation at this shortest wavelength band. The polarization at the longest wavelength band (Band 3) has an azimuthal orientation, which complicates the interpretation of scattering-induced polarization, indicating that additional polarization mechanism(s) may be at work\footnote{Scattering-induced polarization may still be present at Band 3 in HL Tau, but its fraction may be too low (of order $0.3\%$, see Fig.~\ref{fig:longwavpol}e) to compete with other mechanism(s).} \citep{Kataoka2017,Stephens2017,Yang2019}. The lowest spectral index between Band 7 and 6 was found to be $\sim 2$ at the rings, which is indicative of an optically thick ring that may or may not be scattering-dominated. \cite{Carrasco2019} fitted the spectral energy distributions at different radii including both emission and scattering by spherical grains of power-law size distributions and suggested that the relatively low opacity index ($\beta \lesssim 1$) at small radii requires maximum grain sizes in the millimeter range (see their Fig. 7). Such grain sizes would be too large to produce the observed polarization orientation, especially in Band 7, (\citealt{Yang2016_HLTau}, see their Fig. 7). This tension may be alleviated if $\beta$ at small radii is significantly affected by beam averaging, which can average out the high opacity index ($\beta \sim 2$) from the relatively small ($100 \mu$m) grains responsible for scattering-induced polarization in the optically thin gaps and the low apparent opacity index ($\beta \lesssim 0$) from the same (small) grains in optically thick rings (see Fig.~\ref{fig:obsmaj}b for an illustration). Some indication that this may indeed be the case comes from comparing the spectral index between Bands 6 and 7 for the first gap at $\sim 13$~au obtained by \cite{almapartnership2015} at a resolution of $38.6 \times 19.3$ mas ($\beta \sim 1.1-1.5$, see their Fig.~3c) and that by \cite{Carrasco2019} at a resolution of $\sim 50$~mas ($\beta \sim 0.5$, see their Fig.~3, lower panel). It is plausible that at an even higher resolution $\beta$ in the gap may increase further, approaching the standard ISM value of 1.7 for small grains even more closely. There is some hint that the spectral index $\alpha$ near the center is below 2 \citep[see Fig. 2f and 3c of ][]{almapartnership2015}, which provides some additional support to the interpretation of scattering-induced polarization, at least near the center, as in the HD 163296 case; however, the less-than-two $\alpha$ disappears after smoothing with a larger beam \citep{Carrasco2019}. Higher resolution observations and a more detailed analysis of the data are needed to make a firmer case.

\section{Conclusions} \label{sec:conclusion}

In this paper, we have modeled the well-observed protoplanetary disk HD 163296, which has a resolved polarization pattern at ALMA Band 7 ($\lambda=0.87$~mm) strongly indicative of scattering by $\sim 100 \mu$m grains and an apparently low spectral index between ALMA Band 6 ($1.25$~mm) and 7, which is traditionally taken as evidence for emission by large, millimeter-sized, grains. As such, it provides an ideal target to investigate the apparent tension between the grain sizes inferred from polarization and spectral index. Our main results are as follows:

\begin{enumerate}[label=\arabic*)]
    \item We find that our disk model with grains of sizes up to $a_{\rm max}=90 \mu$m broadly reproduces the total intensity (Stokes $I$) observed at $1.25$~mm (ALMA Band 6), the spectral index $\alpha$ between $0.87$ and $1.25$~mm after beam convolution, and the polarization observed at $0.87$~mm (see Fig.~\ref{fig:obsmaj}). In particular, the unusually low spectral index $\alpha \sim 1.5$ observed in the inner disk ($\lesssim 30$~au) is reproduced naturally, as a result of the region being optically thick at both wavebands {\it and} the fact that scattering by the relatively small ($\sim 100 \mu$m) grains is more efficient at the shorter wavelength; the latter flattens the spectral energy distribution, forcing the spectral index below the Rayleigh-Jeans limit $\alpha=2$, as pointed out by \cite{Liu2019} and \cite{Zhu2019}. We are able to reproduce the spectral index inferred for the two innermost rings and gaps as well, but for a different reason. In our model, the rings are optically thick at both wavebands, with an intrinsic spectral index $\alpha < 2$ as a result of scattering (deviation from the  Rayleigh-Jeans power-law also contributes to the low $\alpha$ at the outer ring where the temperature is low). The gaps are optically thin, with an intrinsic $\alpha\sim 4$, as expected for small grains. Convolution with the telescope beam brings the spectral index in the rings up, to $\alpha\sim 2$, and that in the gaps down, to $\alpha\sim 2.5$, both of which are broadly consistent with the observationally inferred values (see Fig.~\ref{fig:obsmaj}). The ability for relatively small, 0.1~mm-sized grains to reproduce the spectral index provides a much-needed, independent piece of evidence for scattering-induced polarization in this source, especially in the inner disk where $\alpha < 2$, which is difficult to understand without scattering. 
    
    \item We infer a midplane temperature in HD163296 that is close to the ALMA Band 6 brightness temperature at the rings. In order for the $\sim 100\ \mu$m-sized grains responsible for scattering-induced polarization to reproduce the observed spectral index of $\alpha\sim 2$, we determine that the rings are mostly optically thick. Our inferred midplane temperature is lower than that inferred from optically thick CO lines. An implication of this is that the latter likely probe the warmer gas from the disk surface down to the so-called ``CO snow surface," but not the colder midplane region where CO is frozen out.
    
    \item Rings and gaps in a disk like HD 163296 produce unique wavelength-dependent patterns in scattering-induced polarization that can be used to further constrain the scattering mechanism. In particular, at shorter wavelengths where the rings are optically thick and gaps optically thin, the polarization fraction is higher in the gaps than in the rings, where the polarization fraction is lowered by large optical depth. Along the gap itself, the polarization is higher along the major axis than along the minor axis (see Fig.~\ref{fig:longwavpol}a). The azimuthal variation is a natural consequence of competition between polarization by radial anisotropy of the radiation field in the gap and polarization by disk inclination. The former is in the azimuthal direction, while the latter is in the direction parallel to the disk minor axis. Together, polarization is enhanced along the disk major axis, but reduced along the disk minor axis. Both the optical depth effect and azimuthal variation are observed in HD 163296 at 0.87~mm (see Fig.~1b of \citealt{Dent2019}). At longer wavelengths where both the rings and gaps are optically thin, the polarization in the rings becomes more prominent relative to that in the gaps, with a higher polarization fraction along the minor axis of the ring (see Fig.~\ref{fig:longwavpol}e); this opposite sense of azimuthal variation is also a consequence of the radiation anisotropy, but in the azimuthal (rather than radial) direction in the rings. Together with polarization induced by disk inclination, polarization is reduced along the major axis, but enhanced along the disk minor axis in the rings. 
    
    \item Scattering-induced polarization has very different spectra across the ALMA Bands (from 0.87 to 3~mm) in rings and gaps. In gaps that are optically thin at all wavebands, the polarization fraction decreases rapidly with increasing wavelength (roughly as $\lambda^{-3}$). In the rings, the polarization may increase with increasing wavelength, depending on the optical depth (see Fig.~\ref{fig:pol_spec}). This inversion of polarization spectrum (i.e., higher polarization at longer wavelength) in optically thick rings but not in the neighboring optically thin gaps is a tell-tale sign of polarization produced by scattering. It should be searched for with high-resolution multi-wavelength polarization observations that resolve the disk substructure.

    \item High-contrast substructures such as rings and gaps now commonly observed in protoplanetary disks complicate the use of spectral index $\alpha$ to infer grain sizes unless they are well resolved. If they are not, the telescope beam averages the low and high values of $\alpha$ in the optically thick and thin regions, respectively, into an apparently intermediate $\alpha$ value that may be misinterpreted as evidence for large, millimeter-sized grains. Accounting for beam averaging has the potential to alleviate the tension between the grain sizes inferred from studies of scattering-induced polarization versus that from spectral index studies, but more work is needed to quantify the extent to which it has affected the grain size estimation in individual sources.
\end{enumerate}

%

\section*{Acknowledgements}
We express our gratitude to W.R.F. Dent for providing the Band 7 Stokes $I$, polarization, and spectral index profiles of HD 163296, L.I. Cleeves for providing the IM Lup Band 7 and Band 6 continuum images and fruitful discussions, and the DSHARP group for making the radial profiles and images of their data publicly available. We also thank C.P. Dullemond for making RADMC-3D publicly available. We thank the referee for constructive comments. Z.D.L. acknowledges support from ALMA SOS. Z.-Y.L. is supported in part by NASA 80NSSC18K1095 and NSF AST-1716259, 1815784, and 1910106. L.W.L. acknowledges support from NSF AST-1910364. C.L.H.H. acknowledges the support of both the NAOJ Fellowship as well as JSPS KAKENHI grant 18K13586. This paper makes use of the following ALMA data: ADS/JAO.ALMA \#2013.1.00226.S, ADS/JAO.ALMA \#2013.1.00601.S, ADS/JAO.ALMA \#2013. 00694.S, ADS/JAO.ALMA \#2015.1.00616.S, ADS/JAO.ALMA \#2016.1.00484.L, ADS/JAO.ALMA \#2016.1.00712.S, and ADS/ JAO.ALMA \#2019.1.00035.S. ALMA is a partnership of ESO (representing its member states), NSF (USA) and NINS(Japan), together with NRC (Canada), MOST and ASIAA (Taiwan), and KASI (Republic of Korea), in cooperation with the Republic of Chile. The Joint ALMA Observatory is operated by ESO, AUI/NRAO and NAOJ. The National Radio Astronomy Observatory is a facility of the National Science Foundation operated under cooperative agreement by Associated Universities, Inc.






\bibliographystyle{mnras}
\bibliography{paper} 



\appendix

\section{An Illustrative Model of A Warmer Disk} \label{app:Tmid}

\begin{figure*}
    \centering
    \includegraphics[width=0.8\textwidth]{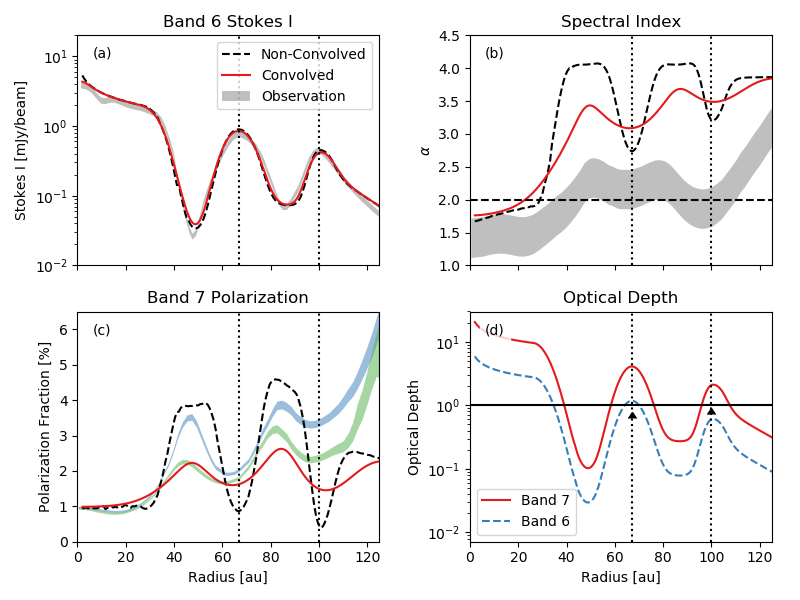} 
    \caption{
        Major axis profiles for the model using a warmer midplane temperature. Panels a-d are plotted in the same way as Fig. \ref{fig:obsmaj}a-d.
            }
    \label{fig:major_compare}
\end{figure*}

To illustrate the effects of a higher temperature, we adopt the temperature profile of \cite{Flaherty2015} (see Fig. \ref{fig:obsmaj}e) and vary the surface density parameters to fit the Band 6 Stokes $I$ high resolution image. The parameters are shown in Table \ref{tab:warmparameters}. The resulting major axis profiles are plotted in Fig. \ref{fig:major_compare}, as in the standard model shown in Fig.~\ref{fig:obsmaj}.

With a higher temperature, the optical depth in Band 6 (Fig. \ref{fig:major_compare}d) has to decrease to reach the same flux. As a result, the B67 and B100 are no longer optically thick, which pushes the spectral index $\alpha$ well above the observational constraint.

We can see in Fig. \ref{fig:major_compare}c that the non-convolved polarization fraction in Band 7 in the gaps for the warmer model is lower than that for the standard model. This is expected since polarization increases with optical depth in the optically thin limit and the optical depth in the gaps for the warmer model is lower than that for the standard model. On the other hand, the polarization in the rings drop to a similar value for both models. The low value of polarization spans a larger region for the standard model, which can be understood since the polarization fraction is roughly constant in the optically thick limit and the rings of the standard model have a larger region where they are optically thick (see Fig. \ref{fig:major_compare}d). After convolution, the polarization fractions from both models appear roughly similar. 

\begin{table}
    \centering
    \begin{tabular}{l c c c c}
        \multicolumn{5}{c}{Parameters} \\
        \hline
        $R_{t}$ [au] & \multicolumn{4}{c}{124} \\
        $T_{t}$ [K] & \multicolumn{4}{c}{22.5} \\
        $q$ & \multicolumn{4}{c}{0.3} \\
        $H_{t}$ [au] & \multicolumn{4}{c}{3} \\
        
        \hlineB{2}\\
    
        \multicolumn{5}{c}{Power-law-like Surface Density} \\
        \hline \\
        Feature             & Inner Disk    & D48   & D86   & B100 \\
        \hline \\
        $\Sigma_{p,0}$ [g cm$^{-2}$]  & 7             & 0.02     & 0.06     & 0.2 \\
        $p$                 & 0.3             & 0     & 0     & 6    \\
        $R_{a}$ [au]        & 0.5             & 38     & 76     & 105 \\
        $R_{b}$ [au]        & 26             & 58     & 96     & 135\\
        $\sigma_{a}$ [au]   & 0             & 5     & 5     & 5 \\
        $\sigma_{b}$ [au]   & 6             & 4.5     & 5     & 10 \\
        
        \hlineB{2} \\
        \multicolumn{5}{c}{Gaussian-like Surface Density} \\
        \hline \\
        Feature             & B67           & B100      & B155\\
        \hline \\
        $\Sigma_{G,0}$ [g cm$^{-2}$]     & 0.9      & 0.3   & 0.04\\
        $R_{0}$ [au]        & 67            & 100   & 155\\
        $\sigma_{c}$ [au]   & 5             & 3     & 5 \\
        $\sigma_{d}$ [au]   & 5             & 4     & 25 \\
        \hline
        
    \end{tabular}
    \caption{
            A chosen set of parameters for the warmer model formatted in the same way as Table \ref{tab:fittedparameters}. 
        }
    \label{tab:warmparameters}
\end{table}

\bsp	
\label{lastpage}
\end{document}